\documentclass[10pt,letterpaper]{article}
\usepackage{opex3}
\usepackage{color}
\usepackage{cite}
\usepackage{amsmath}

\begin{document}

\title{Optimization of multilayered nanotubes for maximal scattering cancellation}

\author{Carlos D\'{\i}az-Avi\~n\'o, Mahin Naserpour, and Carlos J. Zapata-Rodr\'{\i}guez$^{*}$}

\address{Department of Optics and Optometry and Vision Science, University of Valencia, \\ Dr. Moliner 50, Burjassot 46100, Spain}

%

\email{$^*$carlos.zapata@uv.es} 



\begin{abstract}
An optimization for multilayered nanotubes that minimizes the scattering efficiency for a given polarization is derived.
The cylindrical nanocavities have a radially periodic distribution, and the marginal layers that play a crucial role particularly in the presence of nonlocalities are disposed to reduce the scattering efficiency up to two orders of magnitude in comparison with previous proposals. 
The predominant causes leading to such invisibility effect are critically discussed.
A transfer-matrix formalism is additionally developed for the fast estimation of the scattering efficiency of the nanostructures.
\end{abstract}



\bibliographystyle{osajnl} 


\section{Introduction}

The rapid theoretical and experimental advancement in nanomaterials during the last two decades has enabled to engineer multifunctional devices for the control of light with unprecedented proficiency.
For instance, some schemes which are mostly based on transformation optics allow the isolation of a region from interaction with an external light that, in addition, remains unperturbed far from the shadowed space \cite{Leonhardt06,Pendry06,Cai07}.
Consequently, an outside observer cannot detect any target placed inside the cloak by means of an electromagnetic wave field. 
However, such cloaking devices are commonly subject to stringent conditions like the need of materials with exotic electromagnetic parameters reducing their applicability to specific frequencies.
Alternatively we may turn a nanoparticle to be \emph{invisible} by integrating some nanostructured element like ultrathin coatings and metasurfaces in such a way that scattering of the arrangement is much reduced in comparison with the bare object \cite{Alu05,Edwards09,Tricarico09,Filonov12,Kim15}.
In this case, light indeed interacts with the nanocomposite, but scattering from different elements interferes destructively to almost cancel the total scattered signal.

The origin of such scattering drop, thus reducing dramatically the overall visibility of the scatterer, may be attributed to different aspects of the light-matter interaction.
For instance, the local polarizability of distinct components of a moderately sized object with opposite signs may be canceled out in a proper designed configuration  \cite{Alu05,Edwards09,Tricarico09}.
The polarization vector in the elementary materials is anti-parallel with respect to each other, implying that a dipole moment of opposite phase is induced.
Another approach for the cancellation of scattering from an engineered scatterer is based on the properties of the characteristic lineshape of the Fano resonance, where the emission of electromagnetic waves by the object create the interference between the nonresonant scattering from the particle and scattering by narrow Mie modes \cite{Arruda15}.
This effect has also been observed in high-index nanoparticles without additional coating layers \cite{Rybin15}.
Of particular interest follows the inclusion of epsilon-near-zero shells in the spectral range of interest, which may lead to a significant drop of the scattering spectrum and, in addition, create a shielding effect in the bounded space \cite{Filonov12}.
For scatterers with cylindrical symmetry (along the $z$ axis) and composed of dielectric and plasmonic materials, the effect of opposite polarizabilities is mostly observed in TM$^z$-polarized wave fields, whereas Fano resonances may be simply utilized in TE$^z$ polarization configurations.

Recently, a multilayered metallodielectric nanotube with radially-periodic structure has been proposed, which presents a tunable spectral band with significant reduction of its scattering efficiency \cite{Kim15}.
Such scattering cancellation is produced simultaneously for both polarizations and occurs when one of the components of the permittivity tensor characterizing the effective anisotropy of the metamaterial approaches zero, that is near the boundary of the hyperbolic regime \cite{Ferrari15}.
Particularly for TE$^z$-polarized fields, a large birefringence of the nanoshell enables a self-guiding effect along the radial direction that is applied to validate the observed scattering reduction. 

In this work we extend the previous idea of employing radially-periodic metal-dielectric nanotubes to optimize the scattering cancellation of the nanocavity.
For that purpose, we introduce a new degree of freedom that concerns the marginal layers set by the side of the core and the environment medium, the latter playing a critical role primarily in the presence of metamaterial nonlocalities \cite{Diaz16b}.
We also present a critical discussion on the predominant causes leading to such invisibility effect.
The estimation of our results is performed using the full-wave Lorenz-Mie method \cite{Bohren98}, and a matrix formulation is developed in order to further simplify the evaluation of the scattering efficiency,  which presents some similarities to the transfer matrix formalism previously implemented in stratified plane metamaterials \cite{Yeh77}.
In fact, the later formulation might be applied to any complex multilayered cylindrical scatterer composed of homogeneous and isotropic materials.

\section{The transfer matrix formulation}

\begin{figure}[htbp]
 \centering\includegraphics[width=1.0\linewidth]{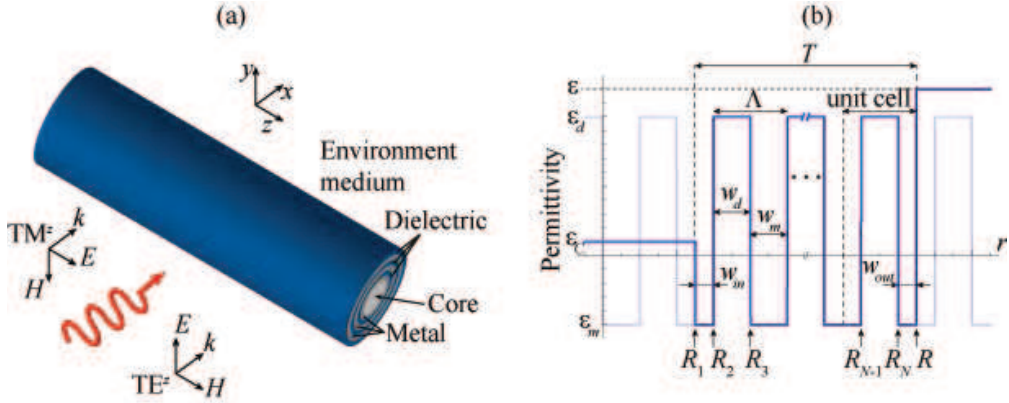}
 \caption{
  (a) Illustration of the coaxial multilayered metamaterial forming the infinitely-long nanotube.
  (b) Permittivity function of our scatterer with radially periodic variation, $\epsilon(r) = \epsilon(r + \Lambda)$, and designer marginal layers.
 }
 \label{fig01}
\end{figure}

Let us consider a cylindrical shell formed by a multilayered metal-dielectric nanostructure, as illustrated in Fig.~\ref{fig01}.
The nanotube, which has a radius $R$ ($= R_{N+1}$), is composed of $N$ layers with a shell thickness given by $T = R_{N+1} - R_1 < R$.
The relative permittivities of the metal $\epsilon_m$ and the dielectric $\epsilon_d$ determine the permittivity distribution of the metamaterial along the radial direction, $\epsilon (r)$.
In this study we consider a periodic distribution, $\epsilon (r) = \epsilon (r + \Lambda)$ within the nanotube, where the period $\Lambda = w_m + w_d$ includes the width of the metal and the width of the dielectric.
In particular, the permittivity $\epsilon_m$ is complex valued, thus taking into account losses in the metal.  
Here we examined nanotubes with a core material of permittivity $\epsilon_C$ and immersed in an environment medium of dielectric constant $\epsilon$.

To estimate analytically the scattering efficiency of the multilayered nanocavity, we followed the Lorenz-Mie scattering method given in detail for instance in Refs.~\cite{Bussey75} and \cite{Shah70}.
Following previously well-established algebraic treatments of plane multilayered photonic structures \cite{Yeh77}, here we propose a transfer-matrix formalism enabling a fast evaluation of the scattering (and potentially extinction) cross section of an infinitely-long cylindrical scatterer composed of a large number of layers. 

In this section we assume that the nanotube is illuminated by a TM$^z$ plane wave propagating along the $x$ axis, as illustrated in Fig.~\ref{fig01}.
Let us point out that the case of TE$^z$ incident wave fields might be simply determined by means of the duality principle \cite{Balanis89}.
The electric field of the incident plane wave may be set as 
\begin{equation}
 \mathbf{E}_{in} = \hat{z} E_0 \exp \left( i k x \right) = \hat{z} E_0 \sum_{n = - \infty}^{+\infty} i^n J_n \left(k r \right) \exp \left( i n \phi \right) ,
 \label{eq03n}
\end{equation}
where $r$ and $\phi$ are the radial and azimuthal cylindrical coordinates, respectively, $E_0$ is a constant amplitude, $J_n(\cdot)$ is the Bessel function of the first kind and order $n$, $k = k_0 \sqrt{\epsilon}$ and $k_0 = \omega/c$ is the wavenumber in the vacuum.
In Eq.~(\ref{eq03n}) we used the Jacobi-Anger expansion of a plane wave in a series of cylindrical waves.
The scattered electric field in the environment medium, $r > R$, may be set as \cite{Bohren98}
\begin{equation}
 \mathbf{E}_{sca} = - \hat{z} E_0 \sum_{n = - \infty}^{+\infty} a_n i^n H_n^{(1)} \left(k r \right) \exp \left( i n \phi \right) ,
 \label{eq06}
\end{equation}
where $H_n^{(1)}(\cdot) = J_n(\cdot) + i Y_n(\cdot)$ is the Hankel function of the first kind and order $n$, and the coefficients $a_n$ must be determined.
The total electric field in the environment medium is simply $\mathbf{E}_{tot} = \mathbf{E}_{in} + \mathbf{E}_{sca}$.
In a given layer of the nanostructured shell (medium $q$), $R_q < r < R_{q+1}$, where $q = \{ 1,2, \cdots, N \}$, the electric field may be expressed analytically as \cite{Bohren98} 
\begin{equation}
 \mathbf{E}_{q} = \hat{z} E_0 \sum_{n = - \infty}^{+\infty} i^n \left[ b_{n,q} J_{n} \left(k_q r \right) + c_{n,q} Y_{n} \left( k_q r \right) \right] \exp \left( i n \phi \right) ,
\end{equation}
where the wavenumber $k_q = k_0 \sqrt{\epsilon_d}$ for a dielectric layer, and $k_q = k_0 \sqrt{\epsilon_m}$ for a metallic layer.
Finally, the electric field in the core of the multilayered tube ($r < R_1$) is expressed as
\begin{equation}
 \mathbf{E}_{C} = \hat{z} E_0 \sum_{n = - \infty}^{+\infty} i^n d_n J_n \left(k_C r \right) \exp \left( i n \phi \right) ,
\end{equation}
where the wavenumber $k_C = k_0 \sqrt{\epsilon_C}$.

The Lorenz-Mie scattering coefficients $a_n$, $b_{n,q}$, $c_{n,q}$, and $d_n$, are determined by means of the proper boundary conditions, that is, continuity of (the $z$-component of) the electric field and the $\phi$-component of the magnetic field, $H_\phi = (i / \omega) \partial_r E_z$, established at the environment-multilayered medium interface given at $r = R$ and at internal interfaces set at $r = R_1$ (boundary with the core) and $r = R_{q+1}$ ($q < N$).
In particular, the boundary conditions applied at $r = R_{q+1}$ may be set in matrix form as
\begin{equation}
 D_{n,q} \left( R_{q+1} \right) \cdot \left[ \begin{array}{c} b_{n,q} \\ c_{n,q} \end{array} \right] 
 =
 D_{n,q+1} \left( R_{q+1} \right) \cdot \left[ \begin{array}{c} b_{n,q+1} \\ c_{n,q+1} \end{array} \right]  ,
\end{equation}
where the \emph{dynamical} matrix
\begin{equation}
 D_{n,m} \left( x \right) = \left[ \begin{array}{cc} J_n \left( k_m x \right) & Y_n \left( k_m x \right) \\ Z_m^{-1} J'_n \left( k_m x \right) & Z_m^{-1} Y'_n \left( k_m x \right)   \end{array} \right]
 \label{eq06b}
\end{equation}
is given in terms of the reduced impedance $Z_q = 1 / \sqrt{\epsilon_m}$ ($Z_q = 1 / \sqrt{\epsilon_d}$) for the metallic (dielectric) $m = q$ layer, and $Z_C = 1 / \sqrt{\epsilon_C}$ for the core, $m=C$.
Here the prime appearing in $J'_n \left( \alpha \right)$ and $Y'_n \left( \alpha \right)$ denotes derivative with respect to the variable $\alpha$.
Note that $D_{n,m}$ is a real-valued matrix provided that the permittivity of the medium $m$ is also real.
By applying the boundary conditions at $r = R$ we may write
\begin{equation}
 D_{n,N} \left( R \right) \cdot \left[ \begin{array}{c} b_{n,N} \\ c_{n,N} \end{array} \right]
 =
 D_{n,N+1} \left( R \right) \cdot \left[ \begin{array}{c} 1 - a_n \\ - i a_n \end{array} \right]  ,
\end{equation}
where, in this case, we set $Z_{N+1} = 1 / \sqrt{\epsilon}$ and $k_{N+1} = k$.

Importantly, we may estimate the fields in the core space and outside the nanotube without calculating the fields in the anisotropic medium by means of the following matrix equation:
\begin{equation}
 \left[ \begin{array}{c} d_n \\ 0 \end{array} \right] =
 M_n \cdot \left[ \begin{array}{c} 1 - a_n \\ - i a_n \end{array} \right]  ,
\end{equation}
where the matrix
\begin{eqnarray}
 M_{n} &=& \left[ \begin{array}{cc} M_{n,11} & M_{n,12} \\ M_{n,21} &  M_{n,22}   \end{array} \right] \\
       &=& \left[ D_{n,C} \left( R_1 \right) \right]^{-1} \cdot \left\{ \prod_{q=1}^{N} D_{n,q} \left( R_{q} \right) \cdot \left[ D_{n,q} \left( R_{q+1} \right) \right]^{-1} \right\} \cdot D_{n,N+1} \left( R \right) . \nonumber
\end{eqnarray}
By using this transfer matrix formalism, it is possible to evaluate analytically the scattering coefficient 
\begin{equation}
 a_n = \frac{M_{n,21}}{M_{n,21} + i M_{n,22}} ,
\end{equation}
which provide the estimation of the scattering efficiency as \cite{Bohren98}
\begin{equation}
 Q_{sca} = \frac{2}{k R} \sum_{n = - \infty}^{+\infty} |a_n|^2 .
 \label{eq07}
\end{equation}
The invisibility condition is established provided that the scattering coefficients $a_n$ (or alternatively $M_{n,21} / M_{n,22}$) arrive simultaneously to a value near zero.


\section{Optimization procedure}

In this study we propose a multilayered metal-dielectric nanotube following a periodic distribution along the radial coordinate.
The basic arrangement consisting of alternating metallic and dielectric layers of widths $w_m$ and $w_d$, respectively, was recently introduced by Kim \emph{et al} in Ref.~\cite{Kim15}.
As a new degree of freedom, we utilize a unit cell within which one of the layers may be surrounded by a second material, and consequently its associated layer must be split into two as illustrated in Fig.~\ref{fig01}(b).
Due to the periodicity of the nanostructure, the internal radial distribution remains unaltered displaying a cycle length $\Lambda = w_m + w_d$, and only the marginal layers may change.
We point out that the marginal layers within a stratified plane nanostructure demonstrate a key role for instance in superlensing \cite{Shamonina01,Feng05,Zapata12a}.
Finally, the integer $T / \Lambda$ provides the number of periods found in the nanotube.

In order to characterize the inmost marginal layer of width $w_{in} = R_2 - R_1$ and the outermost marginal layer of width $w_{out} = R_{N+1} - R_N$, we define the factor $m$ ranging from -1 to +1.
Positive values of $m$ stands for dielectric layers set on the interior and exterior sides of the nanotube.
In this case, by writing $w_{in} = (1 - m) w_d$ and $w_{out} = m w_d$, providing $w_{in} + w_{out} = w_d$, we also include the extreme value $m = +1$ taking into account a full dielectric outermost layer and $m = 0$ referring to a full dielectric inmost layer.
Following an equivalent rule, $m < 0$ is applied to metallic marginal layers, where $w_{in} = -m w_m$ and $w_{out} = (1 + m) w_m$.
In the limit $m = -1$ we consider a full metallic inmost layer, formally the same case accounted for $m = +1$.

In the next numerical simulations we will consider nanotube shells with $T / \Lambda = 3$ periods, that is a total number $N = 7$ layers provided that $0 < |m| < 1$.
We examine silver and TiO$_2$ layers, which permittivities may be analytically approximated within the visible range of frequencies by 
\begin{equation}
\epsilon_m(\lambda) = 3.691 - \frac{9.152^2}{(1.24/\lambda)^2 + i 0.021 (1.24/\lambda)} ,
\label{eq01n}
\end{equation}
and
\begin{equation}
\epsilon_d(\lambda) = 5.193 + \frac{0.244}{\lambda^2 - 0.0803} ,
\label{eq02n}
\end{equation}
respectively \cite{Kim15}.
In the previous equations, the working wavelength $\lambda$ is set in micrometers.

\begin{figure}[htbp]
 \centering\includegraphics[width=.9\linewidth]{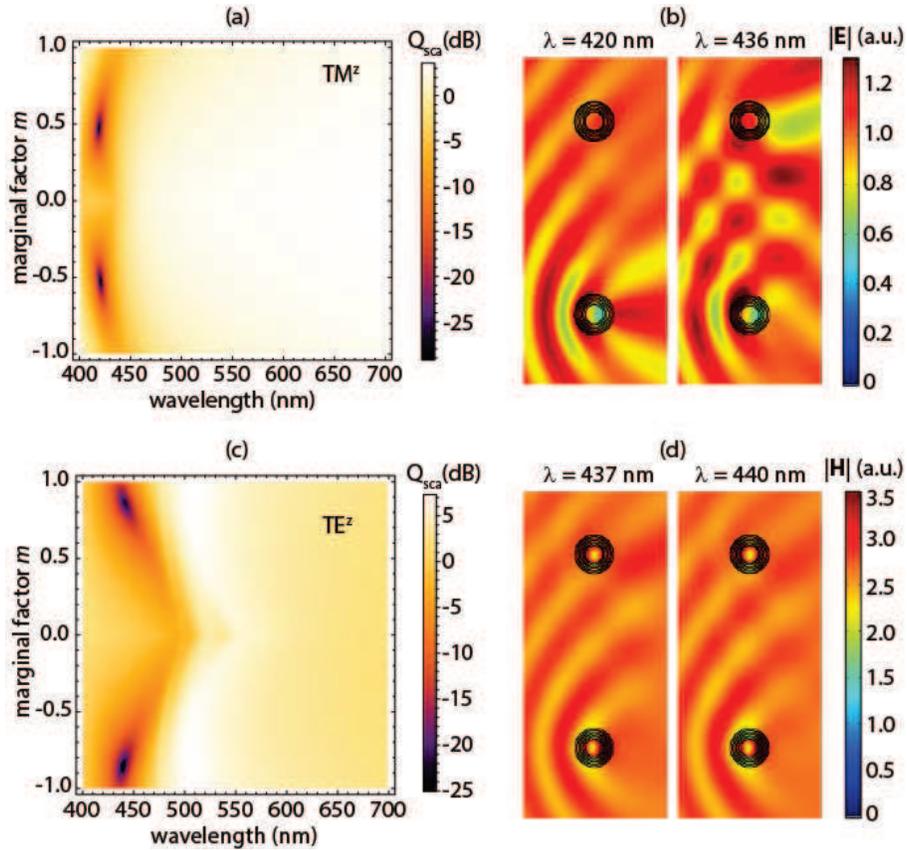}
 \caption{
  Scattering efficiency $Q_{sca}$ of a periodic Ag-TiO$_2$ nanotube immersed in air with inmost radius $R_1 = 50$~nm, thickness $T = 60$~nm, period $\Lambda = 20$~nm, and metal filling fraction $f = 0.5$, when varying the marginal parameter $m$.
  The incident plane wave is: (a)-(b) TM$^z$ polarized, and (c)-(d) TE$^z$ polarized. 
  In (b) and (d) we compare the scattered wave fields of the nanocavities under optimal marginal configurations here analyzed (top scatterer) with the nanotubes proposed in \cite{Kim15} (bottom scatterer) for different wavelengths.
 }
 \label{fig02}
\end{figure}

For the sake of illustration, in Fig.~\ref{fig02} we plot the efficiency spectrum $Q_{sca}$, for wavelengths in the visible, of a nanotube which inmost radius is $R_1 = 50$~nm and thickness $T = 60$~nm.
The scattering efficiency is evaluated using Eq.~(\ref{eq07}) and the transfer-matrix formulation developed above.
For TE$^z$-polarized wave fields, the dynamical matrices given in Eq.~(\ref{eq06b}) should include the transformation $Z_m^{-1} \to Z_m$ in agreement with the duality principle \cite{Balanis89}.
We initially assume a metallic layer width of $w_m = 10$~nm and also a dielectric of width $w_d$ equal to 10~nm, so the metal filling fraction $f = w_m / (w_m + w_d)$ yields 0.5.
The nanotube is immersed in air, where $\epsilon_C = 1$ and also $\epsilon = 1$.
For TM$^z$ polarized wave fields, two local minima in the scattering efficiency are found, as shown in Fig.~\ref{fig02}(a): a first minimum $Q_{sca} = 1.40 \times 10^{-3}$ for the marginal factor $m = - 0.53$ at $\lambda = 421$~nm, in which case the inmost and outermost silver layers approximately have a width of 5~nm each one, and a second minimum $Q_{sca} = 1.20 \times 10^{-3}$ for $m = 0.48$ at $\lambda = 419$~nm, where the inmost and outermost layers are made of titanium dioxide and also have a thickness of around 5~nm.
This represents a substantial improvement of two orders of magnitude with respect to the case $m = -1$ analyzed in Ref.~\cite{Kim15}, where the minimum in efficiency $Q_{sca} = 0.21$ is found at $\lambda = 436$~nm.
Fig.~\ref{fig02}(b) shows the electric field scattered by two nanocavities, one with $m=-0.5$ and the other one with $m=-1$, at the wavelengths of interest.
It is demonstrated that our nanocavity with half-width marginal layers exhibits an extraordinary effect of invisibility ($Q_{sca} = 2.15 \times 10^{-3}$) at the designer wavelength of $\lambda = 420$~nm, and in addition has a similar behavior ($Q_{sca} = 0.27$) compared with the previously proposed nanotube at its best performance wavelength given at $\lambda = 436$~nm. 

In the case that the incident field is a TE$^z$-polarized plane wave, the resultant scattering efficiency of the nanotubes varies considerably as shown in Fig.~\ref{fig02}(c).
It is worth noting the nearly symmetric response of $Q_{sca}$ with respect to the sign of the factor $m$, providing an analogous spectrum when the marginal layers are made of either TiO$_2$ or silver, independently of the modal polarization under consideration. 
Now the minima are found for a marginal factor $m = -0.85$ at $\lambda = 440$~nm, where $Q_{sca} = 3.02 \times 10^{-3}$, and for $m = 0.86$ giving $Q_{sca} = 5.47 \times 10^{-3}$ at $\lambda = 442$~nm.
Note that Kim \emph{et al} reported a scattering cancellation effect giving $Q_{sca} = 5.30 \times 10^{-2}$ at $\lambda = 437$~nm, when we take into account 6 layers and the inmost layer is made of silver ($m = -1$).
Again, our approach provides a remarkably decrease in scattering efficiency, as illustrated in Fig.~\ref{fig02}(d) comparing the wave fields scattered by those two nanocylinders ($m = -0.85$ on the top and $m = -1$ on the bottom) set together.

\section{Discussion}

The interpretation of such significant reduction in scattering may be carried out, at least partially, in terms of the effective medium theory.
For sufficiently narrow slabs ($\Lambda \ll \lambda$), a radially form birefringence may be established for the metamaterial composed of concentric multilayers \cite{Torrent09,Kettunen15}.
In this case, TM$^z$-polarized fields behave like ordinary waves propagating in a uniaxial crystal with optic axis set along the radial coordinate \cite{Yeh88}.
Waves propagate through the metallodielectric metamaterial with negligible variation of the electric field $E_z$, a fact that in addition is in agreement with the electrostatic limit.
However, the electric displacement $\mathbf{D} = D_z \hat{z}$ undergoes critical discontinuities at the metal-dielectric interfaces.
In average, $D_z$ decreases when the metal filling fraction grows, even vanishing in the so-called epsilon-near-zero regime, due to the negative value of the real part of $\epsilon_m$ \cite{Ramakrishna03}.
The $z$-component of the effective permittivity of the layered metamaterial may be estimated accordingly as
\begin{equation}
\langle \epsilon_z \rangle = \frac{\int \epsilon(r) E_z(r,\phi) r \mathrm{d} r \mathrm{d} \phi}{\int E_z(r,\phi) r \mathrm{d} r \mathrm{d} \phi} \approx \frac{2}{\left( R^2 - R_1^2 \right)} \int_{R_1}^R \epsilon (r) r \mathrm{d} r .
\label{eq01b}
\end{equation}
The definition of the weight averaged $\epsilon(r)$ is intuitive, and refined analyses may be found elsewhere \cite{Aspnes13}.
On the other hand, TE$^z$-polarized fields propagate inside the multilayered metamaterial in the same manner as extraordinary waves, where the radial anisotropy is characterized by an average permittivity of components $\langle \epsilon_\phi \rangle$ (that coincides with $\langle \epsilon_z \rangle$) and $\langle \epsilon_r \rangle$, the latter taking much higher values than the former \cite{Kim15}.

Equation~(\ref{eq01b}) yields exactly the well-known expression $\epsilon_\perp = f \epsilon_m + \left( 1 - f \right) \epsilon_d$, associated with the component along the perpendicular direction of the optic axis of the permittivity tensor $\underline{\epsilon}$, in a Cartesian coordinate system, of a form-birefringent metamaterial \cite{Yeh88}, only in the particular case that $m = \pm 1/2$, that is for half-width marginal layers.
Otherwise, the average permittivity $\langle \epsilon_z \rangle$ critically depends on the inmost radius $R_1$ and the marginal parameter $m$; only when the shell is sufficiently narrow, $T \ll R$, enabling the limit $R_1 \to R$, the average permittivity reduces to $\langle \epsilon_z \rangle \to \epsilon_\perp$ in all cases.
On the other hand, further symmetries may be found when the definition of the average permittivity given in Eq.~(\ref{eq01b}) is applied to a metal-dielectric periodic nanocavity.
In particular, $\langle \epsilon_z \rangle$ depends on the absolute value of the marginal parameter $m$.
Since the scattering spectra shown in Fig.~\ref{fig02}(a) and (c) also exhibit such a symmetry, we potentially may establish a correlation of the location of spectral peaks and valleys in terms of the average permittivity.

\begin{figure}[t]
 \centering
 \fbox{\includegraphics[width=\linewidth]{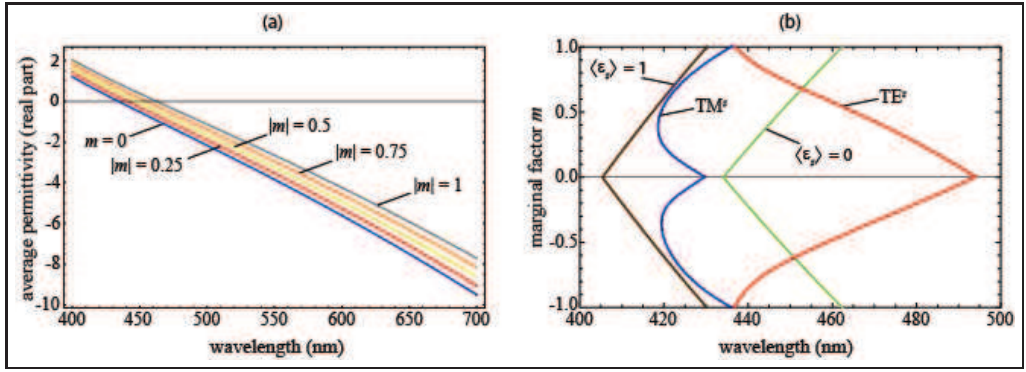}}
 \caption{
  (a) Real part of the average permittivity $\langle \epsilon_z \rangle$ defined in Eq.~(\ref{eq01b}) for an Ag-TiO$_2$ multilayered nanocavity with inmost radius $R_1 = 50$~nm, thickness $T = 60$~nm, period $\Lambda = 20$~nm, and metal filling fraction $f = 0.5$.
  The values of the iso-permittivity curves $\langle \epsilon_z \rangle = 0$ and $\langle \epsilon_z \rangle = 1$ are represented in (b) together with the minima in scattering efficiency found in Fig.~\ref{fig02}(a) for TM$^z$ (blue solid line) and Fig.~\ref{fig02}(c) for TE$^z$ (red solid line).
 }
 \label{fig03}
\end{figure}

In Fig.~\ref{fig03}(a) we plot the real part of the average permittivity $\langle \epsilon_z \rangle$ evaluated in an Ag-TiO$_2$ cylindrical cavity with inmost radius $R_1 = 50$~nm, thickness $T = 60$~nm, period $\Lambda = 20$~nm, metal filling fraction $f = 0.5$, and varying marginal factor $m$.
The average permittivity increases for higher $|m|$ and reaches its maximum value at $|m| = 1$ for the whole spectral range.
For instance, the condition $\mathrm{Re} \langle \epsilon_z \rangle = 1$, where the average permittivity of the multilayered nanotube matches the permittivity of the core and the environment medium, is attained at $\lambda = 405$~nm for $m = 0$ and shifts to longer wavelengths for different marginal factors up to $\lambda = 430$~nm for $|m| = 1$.
As illustrated in Fig.~\ref{fig03}(b), such permittivity matching is behind the invisibility effect observed when the incident plane wave is TM$^z$ polarized, at least for moderate (and high) $|m|$ for which $Q_{sca}$ presents its minima in efficiency.
In the near-optimal case where $m = 0.5$, $\mathrm{Re} \langle \epsilon_z \rangle = 1$ is attained at $\lambda = 417$~nm, very close to the dip shown in Fig.~\ref{fig02}(a) at $\lambda = 419$~nm.
However, the effective medium approximation evidences remarkable limitations.
For instance, there exists critical deviations of the condition $\mathrm{Re} \langle \epsilon_z \rangle = 1$ and the loci of minima in $Q_{sca}$ for low $|m|$. 
More importantly, it cannot predict the extraordinary scattering reduction for marginal factors around $|m| = 0.5$, in comparison with other configurations.

The scene is even more intricate for scattered fields under TE$^z$ polarization.
The minima of the spectral scattering are blue-shifted along with higher values of $|m|$, just advancing to the opposite direction with respect to any iso-permittivity curve with invariant $\langle \epsilon_\phi \rangle$ ($= \langle \epsilon_z \rangle$).
Such severe deviations then put into question the validity of the effective medium theory as an appropriate approach to describe the invisibility effect observed here.
Specifically at the minimum $Q_{sca} = -25.2$~dB found at $\lambda = 440$~nm for $m = -0.85$, the effective permittivity rises to $\mathrm{Re} \langle \epsilon_\phi \rangle = 0.437$.
In practical terms, we might affirm that the invisibility regime is established for marginal factors in the range $0.7 \le |m| \le 1$ occurring in the spectral band that satisfies $0 \le \mathrm{Re} \langle \epsilon_\phi \rangle \le 1$.

Let us point out that, in one side, the existence of a peak near the scattering minimum in every spectra for this specific polarization suggests the occurrence of an isolated Fano resonance \cite{Diaz16c}.
As a consequence of the resonant mechanism of invisibility, it is expected a dramatic impact on external factors such as the permittivity of the core and the environment medium, and also structural aspects such as scale and internal architecture of the nanocavity.
On the other hand, nonlocalities that are inherent in metal-dielectric nanostructures, specially under TE$^z$ polarization \cite{Elser07}, lead for instance to edge effects that seems to be behind an anomalous behavior out of the long-wavelength approximation, as highlighted in Ref.~\cite{Diaz16b}.

\begin{figure}[t]
 \centering
 \fbox{\includegraphics[width=.9\linewidth]{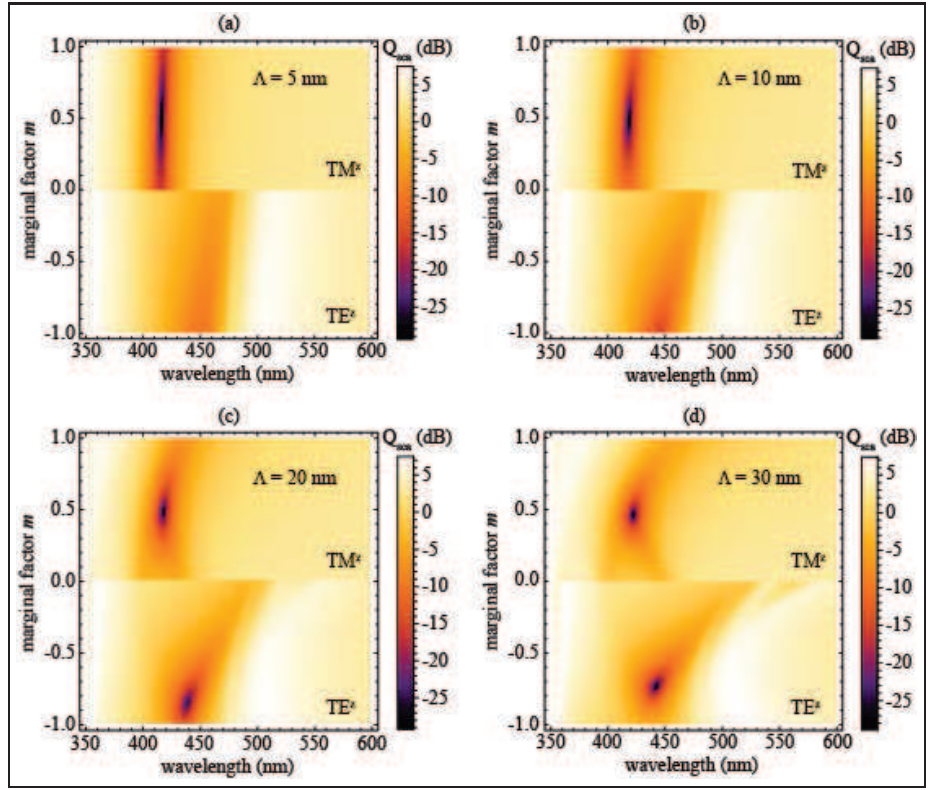}}
 \caption{Scattering efficiency of an Ag-TiO$_2$ nanotube with $R_1 = 50$~nm, $T = 60$~nm, $f = 0.5$, and varying marginal factor $m$, considering a period $\Lambda$ of (a) 5~nm ($N = 25$ layers), (b) 10~nm ($N = 13$), (c) 20~nm ($N = 7$) and (c) 30~nm ($N = 5$). 
  Profiting from the symmetric response of $Q_{xca}$ on the sign of $m$, we represent the scattering spectrum for TM$^z$-polarized waves at $m>0$ and for TE$^z$-polarized fields at negative marginal factors.
 }
 \label{fig07}
\end{figure}

We may obtain further and revelatory conclusions by evaluating the scattering efficiency of our Ag-TiO$_2$ nanotube of $T = 60$~nm for different number of periods.
In Fig.~\ref{fig07} we graphically represent the spectrum of $Q_{sca}$ for nanotubes with $T/\Lambda = 2$, 3, 6 and 12 periods. 
For the smallest period, $\Lambda = 5$~nm, the marginal factor $m$ has a limited influence on the scattering spectrum, as expected; the long-wavelength approximation is valid in this case, and the nanotube might be considered as a radially-anisotropic medium.
For TM$^z$-polarized wave fields, the minimum of scattering efficiency reaches $Q_{sca} = 1.17 \times 10^{-3}$ at $\lambda = 417$~nm.
On the other hand, the scattering efficiency rises significantly when the polarization of the incident plane wave is TE$^z$, for which a minimum value of $Q_{sca} = 8.99 \times 10^{-2}$ at $\lambda = 451$~nm, representing a difference of nearly two orders of magnitude.
Note that such nanostructure embodies a technological challenge since it takes into account a large number of Ag and TiO$_2$ layers, each one with a layer width of 2.5~nm.
When the period increases to $\Lambda = 10$~nm, the impact of the marginal factor is evident as a direct result of nonlocal effects.
In this case, the minimum efficiency for TM$^z$-polarized fields, $Q_{sca} = 1.22 \times 10^{-3}$, is attained at $\lambda = 417$~nm (for $m = 0.5$), manifesting a negligible variation in comparison with the previous case.
For TE$^z$ polarization, the invisibility effect clearly improves in the limiting marginal factor $m = -1$, for which minimum scattering efficiency $Q_{sca} = 2.60 \times 10^{-2}$ got at $\lambda = 443$~nm.
On the other extreme, when the period $\Lambda = 30$~nm, the minima in scattering efficiency are of the same order for both polarizations.
For TM$^z$-polarized fields we have a minimum $Q_{sca} = 1.43 \times 10^{-3}$ at $\lambda = 422$~nm and $m = 0.48$, whereas for TE$^z$ polarization the minimum $Q_{sca} = 1.28 \times 10^{-3}$ is found at $\lambda = 441$~nm for $m = -0.72$, providing a good performance in the first case and the best invisibility behavior in the latter.
In addition, this sort of nanotube with the longest period manifests certain advantages from the practical point of view, such as the reduced number of layers (only 5) with widths of 15~nm except for the margins.
As a consequence, from now on we consider Ag-TiO$_2$ nanotubes of a period $\Lambda = 30$~nm.

\begin{figure}[t]
 \centering
 \fbox{\includegraphics[width=1.0\linewidth]{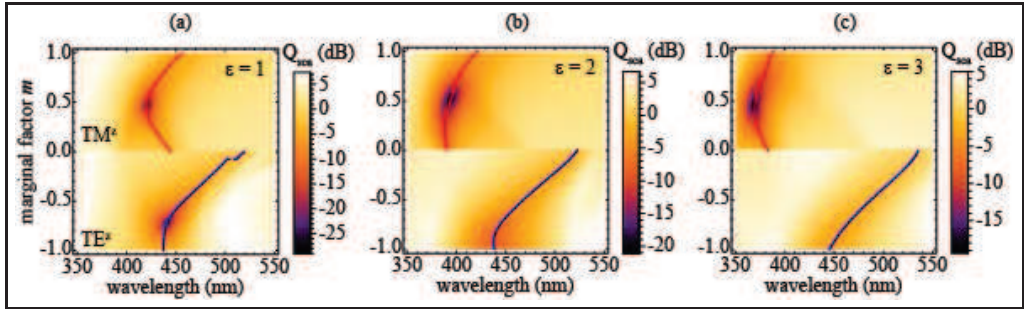}}
 \caption{
  Scattering efficiency of an Ag-TiO$_2$ nanotube with $R_1 = 50$~nm, $T = 60$~nm, $\Lambda = 30$~nm, $f = 0.5$, and varying marginal factor $m$.
  The core and environment medium have a permittivity: (a) $\epsilon_C = \epsilon = 1$, (b) $\epsilon_C = \epsilon = 2$, and (c) $\epsilon_C = \epsilon = 3$.
  We include the loci of minima in $Q_{sca}$ for TM$^z$- and TE$^z$-polarized incident light, colored in red and blue respectively.}
	\label{fig04}
\end{figure}

The influence of cores and environment media with different permittivities is examined in Fig.~\ref{fig04}, for the same Ag-TiO$_2$ nanotube described above.
For brevity we analyzed the case where the dielectric constants $\epsilon_C$ and $\epsilon$ have the same value.
For TM$^z$-polarized incident light, a dramatic drop in the scattering efficiency is again observed but at lower wavelengths when the permittivity $\epsilon$ (and $\epsilon_C$) increases.
In fact, to find the minimum of $Q_{sca}$, the index matching condition may be established as $\mathrm{Re} \langle \epsilon_z \rangle = \epsilon$ in agreement with the shift undergone by the invisibility wavelength at every marginal factor, that is an extremely accurate approach for moderate and high $|m|$.
The lower difference between the permittivity of the dielectric part of the metamaterial and that of the external material, $\epsilon_d$ and $\epsilon$ respectively, requires of a metal with a permittivity also approaching $\epsilon$ which occurs closer to the plasma frequency \cite{Alu05}.
Importantly, the optimal geometrical configuration is found for a marginal factor $|m|$ around 0.5 in all cases.
As an example, a minimum of $Q_{sca} = 1.14 \times 10^{-2}$ is found at $\lambda = 370$~nm for $m = 0.441$ when the permittivity of the environment medium is set as $\epsilon = 3$. 

On the other hand, the scattering efficiency dramatically increases for TE$^z$-polarized wave fields.
The loci of minima in $Q_{sca}$ practically remain in the same spectral band, slightly shifted to longer wavelengths at higher-index core and environment medium, demonstrating a high robustness under changes in the permittivity $\epsilon$.
However, their efficiencies increase several orders of magnitude.
Taking $\epsilon = 3$ for instance, $Q_{sca} = 0.435$ is the minimum efficiency which is found at $\lambda = 449$~nm and $m = -0.935$.
The wavelengths where minima of $Q_{sca}$ are located for TE$^z$-polarized fields seem to be essentially determined by the opto-geometrical characteristics of the nanocavity thus being barely unaltered under changes in the environment; however, this is not longer valid for exceptionally-higher values of the index of refraction and significant deviations might be found  \cite{Diaz16c}. 
This confirms the resonant behavior of the invisibility effect sustained in TE$^z$-polarized wave fields, differently from scattered TM$^z$-polarized waves

\begin{figure}[t]
 \centering
 \fbox{\includegraphics[width=0.55\linewidth]{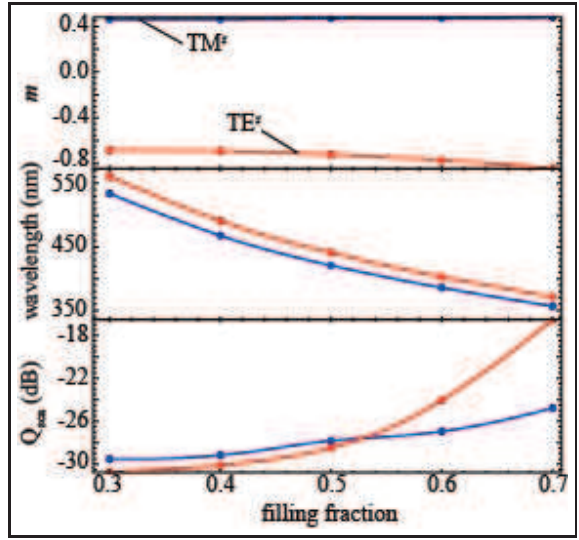}}
 \caption{
  Optimal marginal factor $m$ for invisible nanotubes with $R_1 = 50$~nm, $T = 60$~nm, $\Lambda = 30$~nm, and different metal filling fraction $f$.
  The blue solid line corresponds to nanotubes oriented along the electric field of the incident plane wave, whereas the red solid line refers to TE$^z$-polarized scattered fields.
  We also indicate the minimum scattering efficiency $Q_{sca}$ reached for such optimal configuration, expressed in dB, and the wavelength $\lambda$ for which the latter is achieved.
 }
 \label{fig05}
\end{figure}

In Fig.~\ref{fig05} we consider nanotubes of different metal filling fractions $f = w_m / \Lambda$, but maintaining a nanostructure of period $\Lambda = 30$~nm; again the inmost radius $R_1 = 50$~nm and the nanotube thickness $T = 60$~nm, but the marginal factor $m$ is varied to minimize the scattering efficiency.
A permittivity of unity is set for the core and environment medium in the estimation of $Q_{sca}$. 
The invisibility wavelength in the optimal configuration decreases 200~nm approximately, for both polarizations concurrently, when $f$ varies between 0.3 and 0.7.
This fact suggests a means of tuning the invisibility spectral band of the multilayered scatterers \cite{Kim15}.
For TM$^z$-polarized scattered fields, the undertaken optimization leads to nanocylinders with marginal factors of modulus around 0.5, which provide scattering efficiencies below -25~dB in the range between $f = 0.3$ and 0.7.
On the other hand, the scattering efficiency is reduced to some extent when the incident plane wave is TE$^z$ polarized, at least for the lowest values of $f$.
This most favorable arrangement happens for nanostructures where the inmost silver layer is considerably wider than the outer layer giving marginal factors $m$ neighboring -0.8.
However, $Q_{sca}$ cannot decrease and even reach the limit of -20~dB for the top range of filling fractions, which in principle has no critical impact in the tunability of the invisibility spectral band of the nanotubes expect maybe in ultrasensitive applications.

\begin{figure}[t]
 \centering
 \fbox{\includegraphics[width=1.0\linewidth]{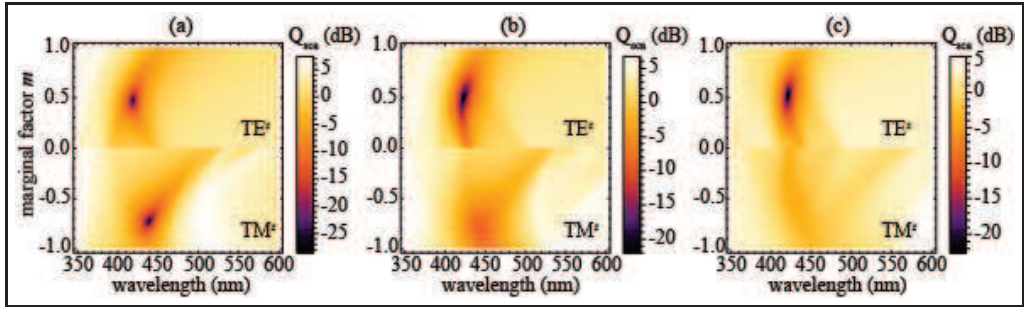}}
 \caption{
  The same as in Fig.~\ref{fig07}(d), but the inmost radius $R_1$ takes values: (a) $50$~nm, (b) $100$~nm, and (c) $200$~nm.
 }
 \label{fig06}
\end{figure}

The size of the metallodielectric nanocavity might be essential concerning the applicability of the effective medium approximation in TM$^z$-polarized wave fields, and more critically the resonant behavior of the nanotube for TE$^z$ polarization. 
In Fig.~\ref{fig06} we plot the scattering efficiency of Ag-TiO$_2$ multilayered nanoshells with different radii immersed in air.
We maintain the scatterer thickness $T = 60$~nm, period $\Lambda = 30$~nm, metal filling fraction $f = 0.5$, and we change the marginal factor $m$ in order to find the configuration with optimal scattering cancellation.
For TM$^z$ polarized fields, the plots remain practically unaltered for inmost radius ranging from $R_1 = 50$~nm up to $R_1 = 200$~nm.
Limited deviations are attributed to the reduced dependence of $\langle \epsilon_z \rangle$ on $R_1$ in spite of the considerable difference in size of the cylindrical scatterers.
Considering now scattered fields under TE$^z$ polarization, the efficiency pattern is modified significantly.
The minimum in scattering efficiency changes from $Q_{sca} = 1.41 \times 10^{-3}$ for $R_1 = 50$~nm as discussed above, increasing to $Q_{sca} = 0.135$ for $R_1 = 100$~nm that is found at $\lambda = 443$~nm when $m = -0.80$, and reaching $Q_{sca} = 0.446$ for $R_1 = 200$~nm (at $\lambda = 426$~nm and $m = -0.43$).
For this specific polarization, the capacity of the nanotubes for canceling the scattered wave field, within the spectral band of interest, decays progressively when the radius $R_1$ grows certainly due to the existence of multiple localized resonances that are associated with whispering-gallery modes.
On the other hand, note that for radius $R_1$ in the micro-scale, such optical microcavities with whispering-gallery modes have stimulated multifunctional applications to optofluidic devices such as microlasers and bio/chemical sensors \cite{Wang14}.

\section{Conclusions}

The study carried out here shows the route for the optimal utilization of the marginal layers in metal-dielectric multilayered nanotubes as invisible conducting scatterers within a tunable spectral band. 
We utilize the inherent nonlocal effects of metallodielectric nanostructures in support of a drastic reduction of the scattered signal of the designer cylindrical nanocavity.  
Our approach leads to a drop of the nanoparticle scattering efficiency that may reach up to two orders of magnitude in comparison with previous proposals also based on radially-periodic arrangements. 
The remarkably invisibility of our Ag-TiO$_2$ nanotube is largely ascribed to scattering cancellation for TM$^z$-polarized incident plane waves and a Fano-type isolated resonance for TE$^z$-polarized fields.
The resonant behavior of the nanotubes in TE$^z$ polarization configurations made it reducing the maximal efficiency to changes in the external environment, diameter of the nanoparticle and even the period of the metamaterial.
Although the cylindrical cavity has a polarization-selective response, it can be adjusted to present a significant reduction of the scattering efficiency simultaneously for both polarizations.
Importantly, the Lorenz-Mie scattering coefficients are set in terms of a transfer matrix formalism leading to the fast evaluation of the scattering efficiency of the multilayered nanocavity.
We believe that the use of invisible conducting nanocavities for invisible electrodes and waveguides presents a promising strategy toward next-generation optofluidic and sensing.

\section*{Acknowledgments}

This work was supported by the Spanish Ministry of Economy and Competitiveness (MINECO) (TEC2014-53727-C2-1-R)

\end{document}